\documentclass[9pt,conference]{IEEEtran}

\usepackage{amsmath,amssymb,amsfonts,amsthm,mathrsfs,mathtools,bbm,stmaryrd}
\usepackage{xparse,pgffor,xspace,glossaries,url}
\usepackage[caption=false,font=normalsize,labelfont=sf,textfont=sf]{subfig}
\usepackage{xcolor}
\usepackage{algorithmic}
\usepackage{algorithm}
\usepackage{cite}
\usepackage{graphicx}
\usepackage{ifthen}
\usepackage{hyperref}

\newcommand{\bs}{\boldsymbol}

\newcommand{\bb}{\mathbb}
\newcommand{\cl}{\mathcal}

\newcommand{\ts}{\textstyle}

\newcommand{\sps}[3]{#3\langle#1,\,#2 #3\rangle}

\newcommand{\cel}{{\sf c}}
\newcommand{\imsymb}{\mathrm{j}}

\newcommand{\GM}[1]{ 
\ifthenelse{\boolean{comments}}
{{\footnotesize \color{blue} [\textbf{GM:} #1]}}
{}
}
\newcommand{\DRAFT}[1]{
\ifthenelse{\boolean{comments}}
{{\color{orange}{#1}}}
{}
}

\newcommand{\casualvar}{\bs z}

\newcommand{\mestim}[1]{\hat{#1}}

\newcommand{\nobs}{Q}
\newcommand{\obsindex}{q}

\newcommand{\obsum}[0]{\sum_{\obsindex = 1}^{\nobs}}

\newcommand{\listobs}[1]{\{#1\}_{\obsindex=1}^\nobs}
\newcommand{\measindex}{m}

\newcommand{\chirpindex}{m_\mathrm{c}}

\newcommand{\matmeas}{\bs Y}

\newcommand{\scatter}{\alpha}

\newcommand{\Noise}{\bs W}

\newcommand{\atom}{\bs a}
\newcommand{\btom}{\bs b}
\newcommand{\atomi}{a}
\newcommand{\btomi}{b}

\newcommand{\rfunsen}{S^\mathrm{r}}
\newcommand{\ufunsen}{S^\mathrm{u}}

\newcommand{\grid}[1]{#1_{\mathrm{G}}}

\newcommand{\domloc}{\cl X}
\newcommand{\domvel}{\cl V}

\newcommand{\locrx}{\bs x^\mathrm{RX}}
\newcommand{\loctx}{\bs x^\mathrm{TX}}

\newcommand{\loc}{\bs x}
\newcommand{\vel}{\bs v}

\newcommand{\range}{r}
\newcommand{\speed}{u}

\newcommand{\carrier}{f_0}
\newcommand{\band}{B}
\newcommand{\tchirp}{T_\mathrm{c}}
\newcommand{\nchirp}{M_\mathrm{c}}

\newcommand{\nsamp}{M_\mathrm{s}}

\newboolean{comments}
\setboolean{comments}{True}

\begin{document}

\newacronym{bomp}{BOMP}{Block Orthogonal Matching Pursuit}
\newacronym{fmcw}{FMCW}{Frequency-Modulated Continuous Wave}
\newacronym{tx}{TX}{transmitter}
\newacronym{rx}{RXs}{receivers}
\newacronym{fft}{FFT}{Fast Fourier Transforms}

\newcommand{\LJc}[2][\footnotesize]{{#1 \color{red}{\bf [LJ: #2]}}}
\newcommand{\GMc}[2][\footnotesize]{{#1 \color{blue}{\bf [GM: #2]}}}
\newcommand{\TFc}[2][\footnotesize]{{#1 \color{magenta}{\bf [TF: #2]}}}
\newcommand{\LVc}[2][\footnotesize]{{#1 \color{green}{\bf [LV: #2]}}}
\newcommand{\MDc}[2][\footnotesize]{{#1 \color{purple}{\bf [MD: #2]}}}
\newcommand{\edt}[1]{\textcolor{blue}{#1}}

\title{\vspace{-3mm} Grid Hopping: Accelerating Direct Estimation Algorithms for Multistatic FMCW Radar}
\author{\IEEEauthorblockN{
        Gilles Monnoyer \IEEEauthorrefmark{1}, 
        Thomas Feuillen\IEEEauthorrefmark{2}, 
        Maxime Drouguet\IEEEauthorrefmark{1}, 
        Laurent Jacques\IEEEauthorrefmark{1}, 
        Luc Vandendorpe\IEEEauthorrefmark{1}
    }
    \IEEEauthorblockA{
        \IEEEauthorrefmark{1} ICTEAM UCLouvain (Belgium).
        \IEEEauthorrefmark{2} SPARC, SnT, UniLu (Luxembourg) \vspace{-3mm}}
}
\maketitle
\begin{abstract} 
    This paper presents a novel signal processing technique, coined grid hopping, as well as an active multistatic \gls{fmcw} radar system designed to evaluate its performance. 
    The design of grid hopping is motivated by two existing estimation algorithms.  
    The first one is the \emph{indirect} algorithm estimating ranges and speeds separately for each received signal, before combining them to obtain location and velocity estimates.
    The second one is the \emph{direct} method jointly processing the received signals to directly estimate target location and velocity.
    While the direct method is known to provide better performance, it is seldom used because of its high computation time.
    Our grid hopping approach, which relies on interpolation strategies, offers a reduced computation time while its performance stays on par with the direct method.
    We validate the efficiency of this technique on actual \gls{fmcw} radar measurements and compare it with other methods.
\end{abstract}
\vspace{-3mm}
\section{Introduction}
\vspace{-.5mm}
Multistatic radar systems are being increasingly used in multiple applications such as low-cost monitoring or automotive systems~\cite{lin2018, saponara2018, capobianco2018}. 
A multistatic radar is composed of one or more \gls{tx} and at least two \gls{rx} with widely spread locations. 
This enables the estimation of the location and velocity of one or multiple targets with an increased diversity by viewing the targets from different angles~\cite{haimovich2008, stinco2014, sun2015}.
In this work, we developed an active K-band \gls{fmcw} multistatic radar 
designed for 
detection of cars in civil applications. 
The focus of this paper is the comparison 
of the performance 
of different estimation algorithms by means of actual measurements. 
Moreover, we propose a novel processing methodology, called grid hopping, which appears as an efficient trade off between the computational complexity and the performance when compared to existing algorithms.

Conventional processing algorithms for multistatic radars can be classified into two categories. 
Indirect methods estimate ranges and speeds of the targets separately for each received signal.
Then they combine the estimates to provide location and velocity estimates through multilateration. 
Direct methods, on the other hand, process all received signals jointly to directly estimate locations and velocities, commonly by maximizing a decision function evaluated over a grid of values~\cite{eldar2010, gogineni2010, berger2011}.
In short, the indirect algorithms are faster as they require the evaluation of a few 2D \gls{fft} that correspond to correlations in the range-Doppler domain.
The direct algorithms require the discretization of the 4D location-velocity domain which makes them 
computationally heavier.
Yet, the results of a direct processing are more robust to non-idealities. 

Grid hopping is inspired by contributions for sound source localization applications~\cite{cobos2011, cobos2017, dietzen2021}.
In brief, we approximate the evaluation of the decision function used by the direct method with an interpolation performed on the output of the 2D \gls{fft} used in the indirect method.
Thereby, the grid hopping enables us to keep the philosophy of the direct method with a computation time similar to the indirect method, and hence provides a compromise between the two strategies.

\textbf{Notations:} Vectors and matrices are denoted by lowercase and uppercase bold letters, respectively. Given a matrix $\bs A$, we use $\bs A^*$, $\bs A^\top$ and $\bs A[i]$ to respectively denote the conjugate, the transpose and the $i$-th row of $\bs A$. The scalar product and the outer product between two vectors respectively read $\sps{\cdot}{\cdot}{}$ and $\otimes$. Finally $\imsymb = \sqrt{-1}$.



\section{Radar Model}

This section introduces the simplified model for the signals acquired by a multistatic \gls{fmcw} radar. 
The model is similar to the one derived in~\cite{monnoyer2019}.
The system uses a modulation made of linear chirps
characterized by a carrier frequency $\carrier$, and bandwidth $\band$ and a chirp duration $\tchirp$. 
A processed frame is composed of the acquisition of $\nchirp$ chirps with $\nsamp$ uniform samples per chirp. 
We restrict here the model to one target.

We consider a multistatic radar composed of one \gls{tx} located in $\loctx$ and $\nobs$ \gls{rx} located in $\listobs{\locrx_\obsindex}$.
The whole system observes a scene that contains a single target assumed to be characterized by a single couple of location-velocity vectors ($\loc$, $\vel$), namely the \emph{parameters of interest}.
We consider subspaces $\domloc$ and $\domvel$ from which $\loc$ and $\vel$ are respectively known to be taken.
Each receiver indexed by $\obsindex$ provides a signal that depends on the range and the radial speed of the target, respectively denoted by $\range_{\obsindex}$ and $\speed_{\obsindex}$.
Those are the \emph{sensed parameters} and are defined with respect to the given receiver  
trough the sensing functions $\rfunsen_{\obsindex}$ and $\ufunsen_{\obsindex}$.
These function are defined as 
\begin{align}
    \range_{\obsindex} 
    & = \rfunsen_\obsindex (\loc) 
     = \ts \frac{2 \band}{\cel} \,\big( \|\loc - \loctx\|_2 + \|\loc - \locrx_\obsindex\|_2\big), 
    \\
    \speed_{\obsindex} 
    & = \ufunsen_\obsindex (\loc, \vel) 
     = \ts \frac{2 f_0 T_c}{\cel} \,\sps{\frac{\loc - \loctx}{\|\loc - \loctx\|_2} + \frac{\loc - \locrx_\obsindex}{\|\loc - \locrx_\obsindex\|_2}}{\vel}{\big}. 
\end{align}

The measurement provided by the $\obsindex$-th \gls{rx} is reshaped into a matrix denoted by $\matmeas_\obsindex \in \bb C^{\nchirp \times \nsamp}$ which can, under a few assumption described in~\cite{bao2014, feuillen2016, monnoyer2019}, be written as
\begin{equation}
    \label{eq:general-model}
    \ts \matmeas_\obsindex = \scatter_{\obsindex} \btom(\speed_\obsindex) \otimes \atom(\range_\obsindex) + \Noise_\obsindex,
\end{equation}
where $\scatter_{\obsindex} \in \bb C$ is a scattering coefficient of amplitude and phase, and where $\Noise_\obsindex$ is a noise term.
The functions $\atom(\range_\obsindex)$ and $\btom(\speed_\obsindex)$ are waveforms or \emph{atoms} whose $\measindex$-th components are respectively given by 
$\atomi_\measindex(\range_{\obsindex}) = \exp({\imsymb 2\pi \,\range_{\obsindex}\, \measindex})$ 
and
$\btomi_\measindex(\speed_\obsindex) = \exp({\imsymb 2\pi \,\speed_{\obsindex}\, \measindex})$.

\section{Direct and Indirect methods}
We summarize now the two classes of existing strategies to recover $\loc$ and $\vel$ from the set of measurements $\listobs{\matmeas_\obsindex}$, namely the indirect and the direct methods.
The indirect method starts by computing the estimates $\mestim\range_\obsindex$ and $\mestim\speed_\obsindex$ independently for each index $\obsindex$. 
This is done by maximizing the range-Doppler map obtained by computing the modulo of the 2D \gls{fft} of $\matmeas_\obsindex$. 
Then, the location $\loc$ is estimated from $\mestim\loc^\mathrm{Ind}$, this quantity being computed by this multilateration
\begin{equation}
    \label{eq:multilat}
    \mestim\loc^\mathrm{Ind} = \ts \arg\,\max_{\loc' \in \domloc} \obsum |\rfunsen_\obsindex(\loc') - \mestim\range_\obsindex|^2.
\end{equation}
The velocity $\vel$ is estimated similarly from $\listobs{\mestim\speed_\obsindex}$ and $\ufunsen_\obsindex$.
Note that the computation time for~\eqref{eq:multilat} is negligible when compared to the computation time required for the $\nobs$ \gls{fft}s computed in the first step.  

The direct algorithm works on grids discretizing the spaces $\domloc$ and $\domvel$ and denoted by respectively $\grid\domloc$ and $\grid\domvel$. 
The factorized methodology for the direct method that is used in~\cite{monnoyer2019} leads to
\begin{equation}
    \label{eq:direct-locestim}
    \mestim\loc^\mathrm{Dir} = \ts \arg\max_{\loc \in \grid\domloc} \obsum \sum_{\chirpindex = 1}^{\nchirp} 
    \big| \sps{\matmeas_\obsindex [\chirpindex]} { \atom(\rfunsen_\obsindex(\loc)) } {\big} \big|^2
\end{equation}
for the direct estimation of the location.
Then, defining ${\bs g}_\obsindex = \matmeas^* \atom(\rfunsen_\obsindex(\mestim\loc^\mathrm{Dir}))$, we compute the velocity estimate as
\begin{equation}
    \label{eq:direct-velestim}
    \mestim\vel^\mathrm{Dir} = \ts \arg\,\max_{\vel \in \grid\domvel} \obsum |\sps{{\bs g}_\obsindex}{\atom(\ufunsen_\obsindex(\mestim\loc^\mathrm{Dir}, \vel))}{}|^2.
\end{equation}

\begin{figure}
    \centering
    \includegraphics[width=.96\linewidth]{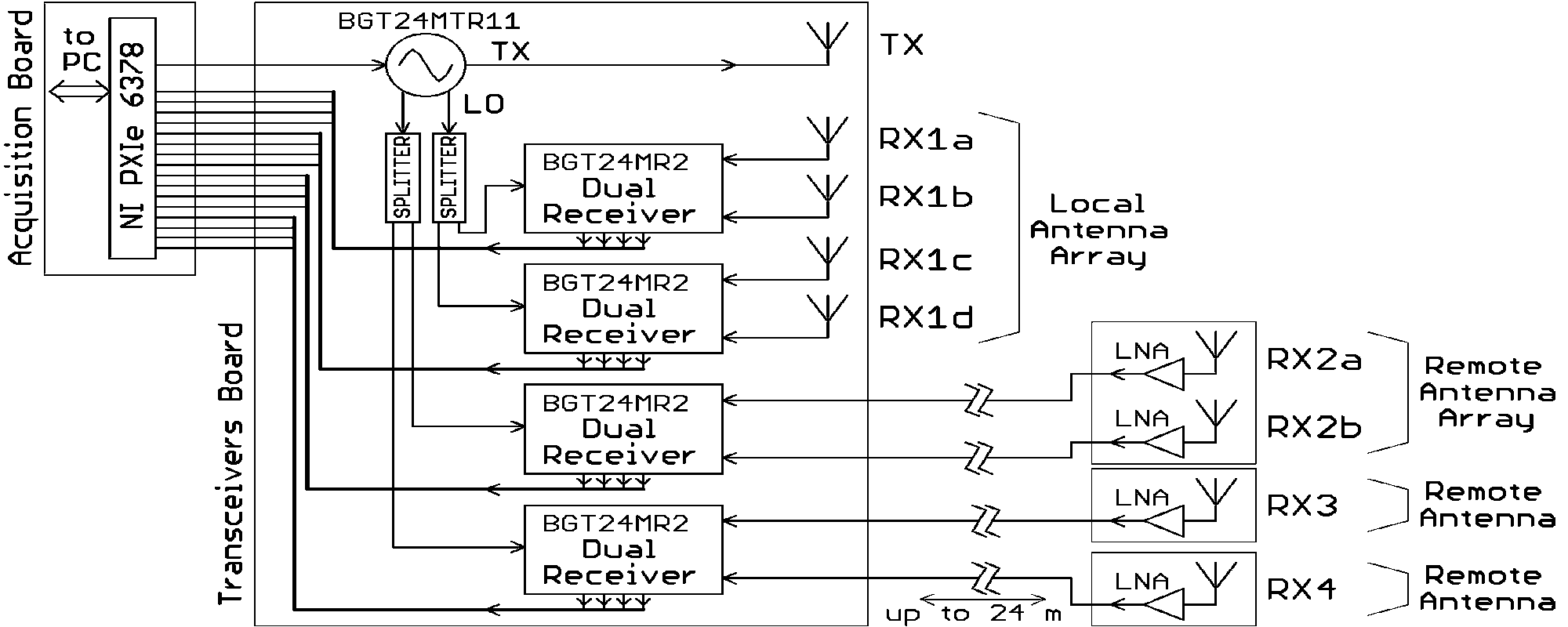}
    \caption{Simplified schematic representation of the hardware.}
    \vspace{-4mm}
    \label{fig:measurement-setup}
\end{figure}

\section{Grid Hopping}
\label{sec:hopping}
Our grid hopping technique relies on an interpolation strategy enabling the approximation of the scalar products on which relies the direct method.
We formulate it for the estimation of location~\eqref{eq:direct-locestim}.
Given a receiver index $\obsindex$  and a column $\chirpindex$,
let us denote by $\casualvar$ the output of the \gls{fft} of the column $\matmeas_\obsindex [\chirpindex]$. 
Then, 
the $i$-th component of $\casualvar$ corresponds to the correlation between $\matmeas_\obsindex [\chirpindex]$ and $\atom(\bar r_i)$ where $\bar r_i$ is the $i$-th range taken from the grid of frequencies (and hence ranges) that is implicitly used by the \gls{fft}.
The grid hopping aims at approximating the scalar products in~\eqref{eq:direct-locestim} as follows.
For all $\loc'\in\grid\domloc$, we identify a set of indexes $\cl I$ and a set of interpolation coefficients $\bs c$ such that 
\begin{equation}
    \label{eq:approx-scalprod}
    \sps{\matmeas_\obsindex [\chirpindex])} { \atom(\rfunsen_\obsindex(\loc')) } {}
    \simeq
    \bs c ^\top \casualvar_{\cl I},
\end{equation}
where $\casualvar_{\cl I}$ denotes the vector $\casualvar$ restricted to the set of indexes $\cl I$.
The quality of the approximation~\eqref{eq:approx-scalprod} is determined by both the interpolation method and the density of the frequency grid used in the \gls{fft} to obtain $\casualvar$.
In this paper, we use the \emph{polar} interpolation whose efficiency has been demonstrated in Fourier atoms interpolation~\cite{simoncelli2011, duarte2013, duarte2014}.
The computation of the interpolation coefficients for all $\loc'\in\grid\domloc$ is heavy but performed offline.

To apply grid hopping to the velocity estimation~\eqref{eq:direct-velestim}, we cannot compute the index sets and the coefficients offline; the speed sensing function $\ufunsen$ 
indeed depends on the location that must be first estimated.
To avoid the online computation of interpolation coefficients, we use for the velocity estimation the simplest interpolation scheme where each set $\cl I$ contains a single index 
for all $\vel'\in\grid\domvel$.

To summarize, grid hopping follows the same principle as the direct method.
Yet, the explicit scalar products required in~\eqref{eq:direct-locestim} and \eqref{eq:direct-velestim} are replaced by an interpolation procedure such as~\eqref{eq:approx-scalprod}. 
Indirect, direct and grid hopping strategies are compared hereafter.

\begin{figure}[tb!]
    \centering
    \includegraphics[width=.52\linewidth]{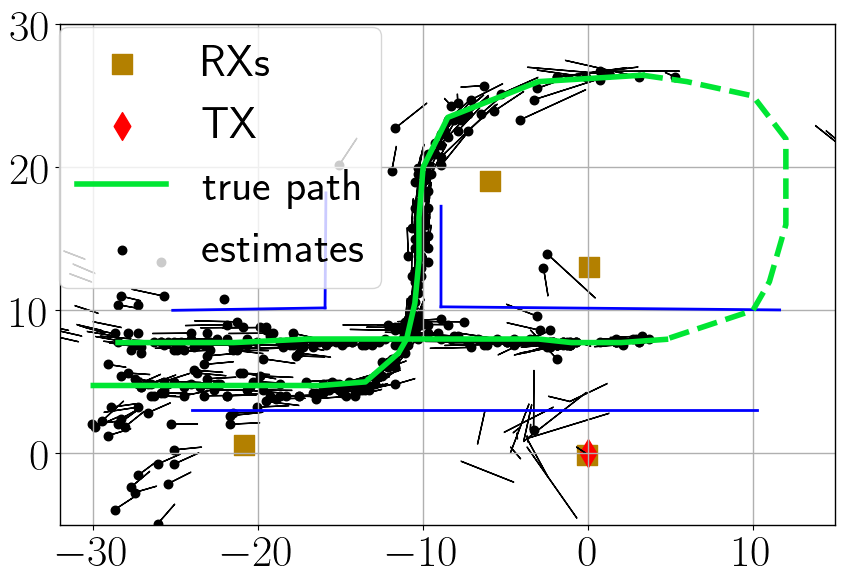}
    \includegraphics[width=.46\linewidth]{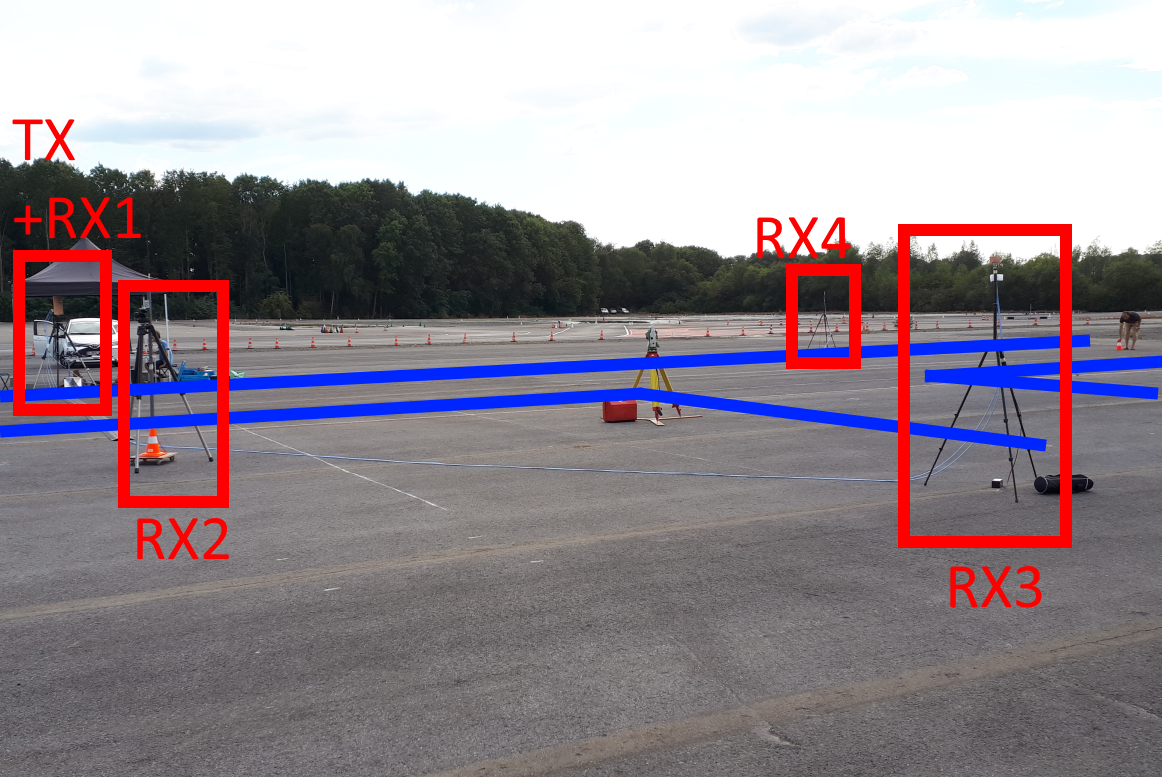}
    \caption{\textbf{(left)}
    Geometry of the system, showing the locations of the TX and the RXs. 
    The axis depicts the 2D ground in meters.
    The dashed part of the car path is a portion that we discard from the measurement because the car was outside the sight of all receiving antennas. 
    The dots and arrows are respectively the location and velocity estimates resulting from the hopping algorithms with grid densities of half the resolution.
    \textbf{(right)} Photo of the actual setup.
    }
    \label{fig:setup}
\end{figure}

\begin{figure}[tb!]
    \centering
    \vspace{-4mm}
    \includegraphics[width=.45\linewidth]{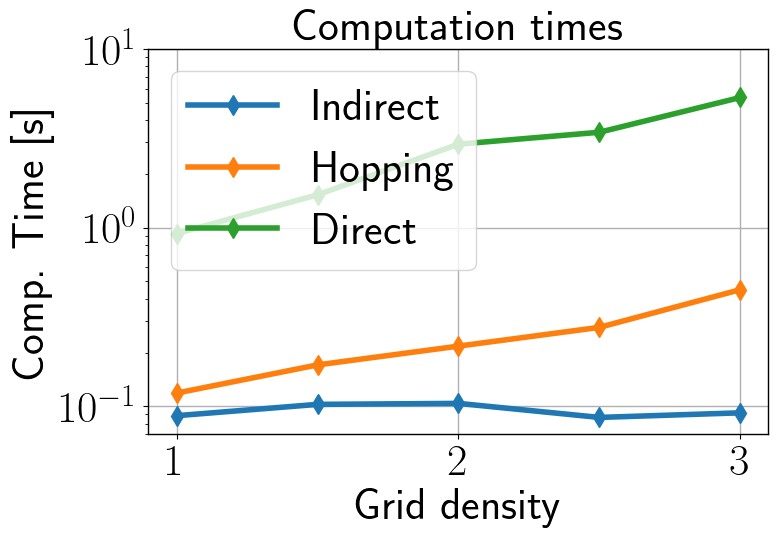}
    \includegraphics[width=.45\linewidth]{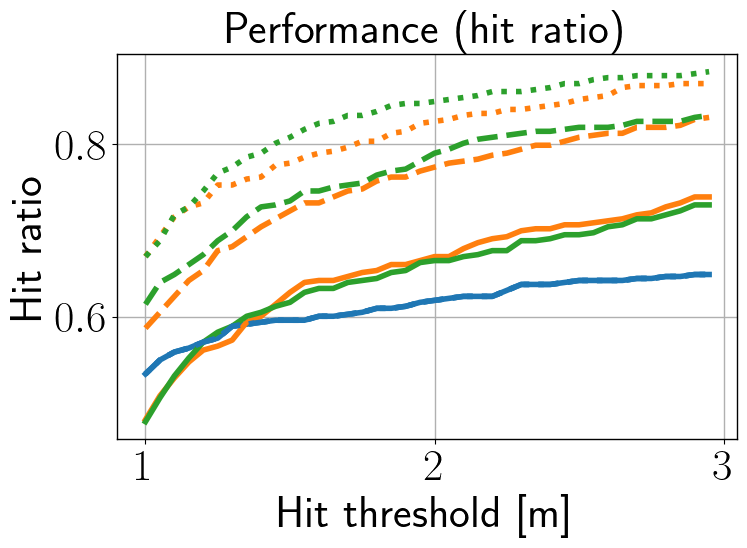}
    \caption{
    Computation time and performance comparison of the three algorithms.
    The hit ratio is the amount of hits divided by the total number of processed frames (433).
    (In right graph) Solid line : grid density $d=1$. Dashed : $d=2$, Dotted : $d=4$.
    A grid density $d$ means that the spacing between adjacent bins of $\grid\domloc$ is $1/d$ times the radar's range resolution (60cm).
    }
    \vspace{-4mm}
    \label{fig:results}
\end{figure}

\section{Experimental results}
To compare the three algorithms, 
we designed an active multistatic radar system and measured different scenarios with one to four cars moving along controlled patterns.
The radar system is built around the BGT24 family of radio frequency transceivers from Infineon. Figure \ref{fig:measurement-setup} shows a simplified schematic of the hardware. 
The \emph{base station} includes a computer, the acquisition board, one transmitter and a four antennas array.
Three \emph{remote stations} are connected to the base station through coaxial cables of length up to 24 m, and are equipped with respectively a two antennas array, twice a single antenna. 
A general purpose data acquisition board (NI 6378 from National Instruments) outputs voltage ramps toward the voltage controlled oscillator (BGT24MTR11) to generate chirps of duration $\tchirp = 128\mu$s around $\carrier= 24 GHz$, with $\band = 250MHz$. 
This signal feeds the \gls{tx} antenna and is split to also feed all the \gls{rx}. The receivers are IQ demodulators (BGT24MR2) followed by baseband amplification and filtering circuits. The remote stations have a local low noise amplifier to compensate the losses in the cables. The baseband signals are sampled by the acquisition board such that $\nsamp = \nchirp = 128$ and sent to the computer for processing. 

This paper presents a first glimpse of the results. 
We focus on a scenario represented in Fig.~\ref{fig:setup} where a single car is moving along the path depicted in the figure. 
For simplicity, one RX antenna in each station was used for the processing presented in this paper.
The radar signals were acquired for 60 seconds at a rate of one processed frame of $\nchirp \nsamp$ samples every 60ms on average. 
After discarding the frames where the car was outside the sight of all \gls{rx}, $433$ processed frames are left to assess the performances of the different algorithms. 

The performance and the computation time of the three methods are compared in Fig.~\ref{fig:results}. 
Each location estimate resulting from each algorithm was compared to the approximated
ground truth drawn in Fig.~\ref{fig:setup}. 
As the car is not a single point in space, such comparison can only make sense with a tolerance threshold. 
When the error between an estimate and the reference is smaller than the ``hit threshold", we counted it as a hit (see Fig.~\ref{fig:results}).
The performance was tested for multiple densities of $\grid\domloc$.
We observe that the direct method performs better than the indirect one with a higher computational time.
The hopping strategy provides an alternative with a computation time closer to the indirect method while exhibiting performance similar to the direct method. 

\section{Conclusion}
In this paper, we presented a novel methodology, called grid hopping, as a trade-off between the direct and the indirect methods.
The efficiency of grid hopping was evaluated on actual radar measurements. 
The comparison was restricted to a single interpolation scheme and to our most simple scenario with a single car moving. 
In the future, we will study the grid hopping to iterative direct algorithms 
enabling the detection of multiple targets.
We will also compare multiple interpolation schemes.

\bibliographystyle{IEEEtran}
\bibliography{refs/ref-radar,refs/ref-sl,refs/ref-sparse}
\end{document}